# Large Magnetoresistance in Topological Insulator Candidate TaSe$_3$

*Yong Zhang*[1,2,†], *Tongshuai Zhu*[1,†], *Haijun Bu*[1], *Zixiu Cai*[3], *Chuanying Xi*[4], *Bo Chen*[1,2], *Boyuan Wei*[1,2], *Dongjing Lin*[1], *Hangkai Xie*[1,2], *Muhammad Naveed*[1,2], *Xiaoxiang Xi*[1], *Fucong Fei*[1,2]\*, *Haijun Zhang*[1]\*, *Fengqi Song*[1,2]\*

[1] *National Laboratory of Solid State Microstructures, Collaborative Innovation Center of Advanced Microstructures, and School of Physics, Nanjing University, Nanjing210093, China.*

[2] *Atomic Manufacture Institute (AMI), Nanjing 211805, China.*

[3] *School of Physics, University of Wisconsin-Madison, Madison 53706, Wisconsin, USA.*

[4] *High Magnetic Field Laboratory, Chinese Academy of Science, Hefei 230031, Anhui, China.*

[†]These authors contributed equally to this work. *Author to whom correspondence should be addressed: feifucong@nju.edu.cn; zhanghj@nju.edu.cn; songfengqi@nju.edu.cn.

**Abstract**

Large unsaturated magnetoresistance (XMR) with magnitude $\sim 10^3$% is observed in topological insulator candidate TaSe$_3$ from our high field (up to 38 T) measurements. Two oscillation modes, associated with one hole pocket and two electron pockets in the bulk, respectively, are detected from our Shubnikov-de Hass (SdH) measurements, consistent with our first-principles calculations. With the detailed Hall measurements performed, our two-band model analysis exhibits an imperfect density ratio $n_h/n_e \approx$ 0.9 at *T*< 20 K , which suggests that the carrier compensations account for the XMR in TaSe$_3$.

## I. INTRODUCTION

Recent years has witnessed the great development of the topological materials[1-6] and its quantum devices[7-11] for their exotic transport properties[12-14]. One of the most studied phenomena is the large unsaturated magnetoresistance (XMR)[15,16]. The common existence of XMR has been reported in topological semimetals, including $Cd_3As_2$[1], $WTe_2$[16-19], $MoTe_2$[20,21], TaAs family[2,22-25], lanthanum monopnictide family[26-31], etc.

Nevertheless, XMR has yet been found in any topological insulators, such as $Bi_2Se_3$ family[32], $PbSn_{1-x}Se_x$[33], BiTeI[34-36]. It is not hard to understand because the bulk states remains insulating at the Fermi surface for a typical topological insulator[37]. Interestingly, a kind of material named $TaSe_3$ with bulk metallic states was recently predicted to be a strong topological insulator candidate [38]. It is a quasi-one-dimensional trichalcogenide with space group $P2_1/m$ (No. 11) and the crystal structure are shown in Fig. 1(a). In previous works, single crystal $TaSe_3$ yields a residual resistance ratio (RRR) around 20 [39] and $TaSe_3$ nanowires claims a low resistivity and high breakdown current density[40-42]. In addition, $TaSe_3$ mesowires possibly host a charge density wave (CDW) at 65 K[43].

Here, we systematically investigate the electronic structure of $TaSe_3$ and its corresponding transport behavior in this material by combining the first-principles calculations and low-temperature and high-field transport measurements. Unsaturated XMR with magnitude $\sim 10^3$% is observed from our high field (up to 38 T) measurements. Besides, two oscillation modes, associated with one hole pocket and two electron pockets in the bulk, respectively, are detected from our SdH measurements, consistent with our first-principles calculations. Moreover, our two-band model analysis exhibits that an imperfect density ratio $n_h/n_e \approx 0.9$ accounts for XMR when $T<$ 20 K, based on our detailed Hall measurements data.

## II. METHODS AND RESULTS

To experimentally investigate the transport properties of this topological insulator candidate TaSe$_3$, single crystals were synthesized via chemical vapor transport (CVT) method. Ta, Se powders were mixed in a ratio of 1:6 where the Se itself played the role of transport agent. The mixture was vacuum sealed in a quartz tube, and then heated up to high temperature with a gradient from 1000 °C to 600 °C in a two-zone furnace and kept for one week. Then the furnace was naturally cooled down and the ribbon-like crystals were obtained with typical dimensions of 5 mm × 0.1 mm × 0.05 mm, where the longest dimension corresponds to crystallographic *b* axis (along the Ta-Ta chains). The chemical components of the obtained TaSe$_3$ crystals were identified by the energy dispersive spectrum (EDS), as shown in Fig. 1(b). An atomic ratio Ta : Se = 1 : 2.73 could be extracted from the EDS, suggesting a slight amount of Se vacancies in our samples. The insets of Fig. 1(b) show the EDS mapping of Ta and Se, indicating the uniform distribution of the two elements in the obtained TaSe$_3$ samples. Raman measurements with 532 nm laser excitation in the backscattering geometry (Princeton instruments) were also performed. Multiple peaks have been observed in Raman measurements, as shown in Fig. 1(c), which is consistent with the previous study[41,43,44]. We measured the temperature dependence of resistance of TaSe$_3$ samples, as displayed in Fig. 1(d). The resistance decreases when cooling down, displaying a metallic behavior with residual resistivity ratio (RRR)$R(300 \text{ K})/R(2 \text{ K}) = 23$, which is similar to the literature[39].

### A. XMR AND QUANTUM OSCILLATIONS

We then carried out the magneto-transport measurement of our samples under high magnetic field up to 38 T. Fig. 2(a) displays the field dependence of magneto-resistance (MR) of TaSe$_3$ with the current along the b axis and the applied field along the [1 0 -1] direction at various temperatures from 0.36 K to 78 K. The MR is defined by $[R_{xx}(B) - R_{xx}(0)]/R_{xx}(0) \times 100\%$. It is noticeable that the MR reaches 1800% at 0.36 K and keeps continuously increasing when raising the magnetic field, indicating a non-saturated XMR behavior. In order to further quantitatively analyzing the oscillation behavior, a smooth background was subtracted from the longitudinal resistance $R_{xx}$ and the oscillation patterns as a function of $1/B$ at representative temperatures are presented as seen in Fig. 2(b).

By fast Fourier transformation (FFT) based on the oscillation components displayed in Fig. 2(b), two major peaks with frequency of $F_\alpha = 99$ T and $F_\beta = 173$ T are uncovered, as shown in Fig.2(c). In general, the SdH oscillations can be well described by the Lifshitz-Kosevich (LK) formula[45],

$$\Delta R_{xx} \propto R_T R_D \cos\left[2\pi\left(\frac{F}{B} - \frac{1}{2} + \varphi\right)\right]. \tag{1}$$

The thermal damping factor is $R_T = \frac{\chi T}{\sinh(\chi T)}$, and the Dingle damping factor is $R_D = \exp(-\chi T_D)$, where $\chi = \frac{2\pi^2 k_B m^*}{\hbar e B}$. $\varphi$ is the phase shift, and $\varphi = \frac{\varphi_B}{2\pi} - \delta$, where $\varphi_B$ is the Berry phase and $\delta$ equals 0 and $\pm 1/8$ for two-dimensional and three-dimensional systems, respectively. $k_B$ denotes the Boltzmann constant; $\hbar$, the Planck's constant; $F$, the frequency of the oscillation mode; $m^*$, the effective mass. The effective mass $m^*$ can be extracted by fitting the temperature dependence of the corresponding oscillation amplitude to $R_T$, as shown in Fig. 2(d), having $m^*_\alpha = 0.63\, m_e$ and $m^*_\beta = 0.75\, m_e$. Since there are two oscillation modes in this material, the other related parameters, such as Dingle temperatures, quantum mobilities, etc., need to be extracted by the multiband LK formula[46,47], as depicted in Fig. 2(f). In Fig. 2(e), we obtain the intercepts at $n$ axis of -0.03 and -0.02, respectively, which are both near to zero, indicating the zero Berry phases. These results are the same as the ones got from LK fitting process as seen in Table S2 (see supplementary materials). Note that the ratio $\rho_{xx}/\rho_{xy} \approx 10$ to 100 from Hall measurements under a relatively low field, so the longitudinal conductance could be estimated as $\sigma_{xx} \sim \frac{1}{\rho_{xx}}$, where the peaks of $\rho_{xx}$ should be located at the integers of the Landau fan diagram. Comparing to our data, one recent work came out with different Berry phase values, which might arise from the lack of Hall data, not noticing the ratio $\rho_{xx}/\rho_{xy}$ is relatively large[48]. The extracted almost zero Berry phases indicates that these Fermi pockets in TaSe$_3$ should belong to bulk pockets rather than the topological surface states.

## B. FIRST-PRINCIPLES CALCULATIONS

In order to obtain the basic understandings of the topological nature of electronic structure as well as the Fermi pockets of TaSe$_3$, we performed the first-principles calculations.

Fig. 3(a) displays the calculated band structure without spin-orbit coupling (SOC), noticing that a band crossing can be identified as marked by the dashed rectangle. The crossing points are gapped when inducing SOC as displayed in Fig. 3(b). By calculating the topological invariant $Z_2$ through the Wannier charge center(WCC)[49], TaSe$_3$ proves to be a strong topological insulator since the WCC of six time-reversal invariant momentum plane (Fig. 3(e)) guarantee that $Z_2$ = (1; 100),consistent with the previous report[38].Fig. 3(d) shows the surface states on (0 1 0) plane of TaSe$_3$, where one Dirac cone of topological surface states lies at $\bar{B}$ point. From the view of topology classification, TaSe$_3$ is really a strong topological insulator. However, the strong topological insulator state is defined based on a so-called curved Fermi level, which means no band gap actually[38].

To obtain the correct band structure, we made a quantitative comparison between the calculated extremal areas of the carriers' pockets projecting on (1 0 -1) plane and the experimental one. On one hand, we calculated the band structure with the crystal structure parameters from ICSD database in the first place, but the calculated results were 2 times larger than the experimental ones. On the other hand, in the recently reported reference[38], it tells that the a-axis and the c-axis parameter may be affected by the temperature or the strain easily, as TaSe3 is a layered structure with trigonal-prismatic chains going in the b direction. Thus, we think that the affected crystal structure may account for the mismatching between the calculations and experiments. By adjusting the crystal structure parameter a little bit, a self-consistent calculated result obtained (see Fig. S2 in supplementary material). According to the Onsager relation $S_f = (2\pi e/\hbar)F^{50}$, the extremal cross-sectional areas are $0.94 \times 10^{-2}$Å$^{-2}$ and $1.69 \times 10^{-2}$Å$^{-2}$ for the quantum oscillation frequencies $\alpha$ (99 T) and $\beta$ (173 T), respectively. Fig. 3(f) shows the Fermi surfaces of TaSe$_3$. With marking the cut parallel with (1 0 -1) plane through gamma point as $k_{(1\,0\,-1)} = 0$, We located the extremal cross-sectional area of the hole pocket enclosing the gamma point with an area of $1.86 \times 10^{-2}$Å$^{-2}$ at $k_{(1\,0\,-1)} = 0$, as shown Fig. 3(g), while that of one of the two identical electron pockets across the boundary of the first Brillouin zone (BZ) with an area of $0.91 \times 10^{-2}$Å$^{-2}$ at $k_{(1\,0\,-1)} = -0.46\pi$, as shown in Fig. 3(h). Thus, we conclude that the $\alpha$ frequency corresponds to the two identical electron pockets crossing the boundary of the first BZ and the $\beta$ frequency to the hole one enclosing the gamma point.

## C. HALL MEASUREMENTS

Next, we performed the detailed Hall measurements on this compound under a relatively low field up to 9 T. Fig. 4(a) and 4(b) display the plots of longitudinal resistivity $\rho_{xx}$ and Hall resistivity $\rho_{xy}$ as functions of $B$ at different temperatures ranging from 2 K to 50 K. It is worth noticing that a sign change of $\rho_{xy}$ at low field range from positive to negative occurs when increasing the temperature from 15 K to 50 K (precisely, 20 K to 30 K), which may account for the sudden change in carriers density and mobility in Fig. 4(e) and Fig. 4(f) around at 20 K, due to the possible crystal structure parameters affected by the temperature[38]. Besides, the nonlinear behavior of $\rho_{xy}$ gives the indication for electrical conduction by both electron and hole carriers.

Thus, the Hall response was then further analyzed by the isotropic two-band model, which was usually used to identify the transport contributions from hole-like and electron-like carriers[20,51,52]:

$$\sigma_{xx} = \frac{en_e\mu_e}{1+(\mu_e B)^2} + \frac{en_h\mu_h}{1+(\mu_e B)^2}, \quad (2)$$

$$\sigma_{xy} = -\sigma_{yx} = \left[\frac{n_e\mu_e^2}{1+(\mu_e B)^2} - \frac{n_h\mu_h^2}{1+(\mu_h B)^2}\right]eB. \quad (3)$$

The results of $\sigma_{xx}$ and $\sigma_{xy}$ based on Eqs. (2) and (3) are shown in Figs. 4(c) and 4(d), respectively, plotted in the form of solid curves, matching with the experimental data except a tiny diverge where field closes zero for $\sigma_{xx}$. Considering the fact that the Hall signals are more sensitive to the carriers near the Fermi surface, we plot the detailed extracted parameters, densities of carriers $n_e$ and $n_h$ and mobilities of carriers $\mu_e$ and $\mu_h$, from the two-band model fittings of $\sigma_{xy}$ as a functions of temperatures from 2 K to 60 K, as shown in Figs. 4(e) and 4(f). Interestingly, we find that the ratio of densities of carriers $n_h/n_e$, as shown in Fig. 4(e), stays approximately 0.9 when temperature less than 20 K. The imperfect but balanced enough ratio of carriers could cause considerable compensation between electrons and holes, which may account for the non-saturated XMR at low temperature in TaSe$_3$. In previous theoretical calculations, the ratio of two types of carriers $n_h/n_e \approx 1$ will lead to a unsaturated magnetoresistance forever, like in WTe$_2$[4,16-19]; in our case, the ratio $n_h/n_e \approx 0.9$ can result in a XMR reaching the magnitude of $10^4$ % before saturation[53]. Thus, it is understandable to attribute the non-saturated MR behavior to the compensation of two types of carriers, even for the materials with distinct band topology. The difference of the magnitude between our data and the theoretical estimation is probably due to the low mobility of carriers[26,54].

### III. DISCUSSION AND SUMMARY

In summary, we carried out systematic investigation on the electronic structure and the corresponding transport behavior for TaSe$_3$ by combining the first-principles calculations and low-temperature and high-field transport measurements. With the analysis of two-band model, based on our detailed Hall measurements data, we reveal that the origin of the XMR may be associated with the carrier compensations in TaSe$_3$.

Unfortunately, only bulk properties are detected and the signals from the topological surface states are still concealed, unlike the recently reported work[48] claims revealed, at least from the perspective of the transport measurements. There are a few possible reasons for the absence of the topological surface states signal of TaSe$_3$ from our experiments. Firstly, there is no real band gap in this system, thus the transport property is dominated by the bulk carriers and the metallic behavior has been confirmed by our transport measurements, while the contribution from the surface states may be masked over. This may put TaSe3 as a semimetal with topological surface state, the same category as the one that the lanthanum monopnictide family [26-31] is in. Secondly, the calculated topological surface states on (1 0 -1) plane [38], where the field are perpendicular to, is rather subtler than that on (0 1 0) plane and quite lower to the Fermi level, making difficulties to detect. Though the smoking gun to experimentally confirm the existence of topological surface states still need to be verified by further studies. Our work provides an opportunity to exploit the interplay between the topology and unsaturated XMR in a topological insulator.

**SUPPLEMENTARY MATERIAL**

See the supplementary material for detailed refined calculations and fitted parameters from SdH oscillations.

**ACKNOWLEDGEMENTS**

The authors gratefully acknowledge the financial support of the National Key R&D Program of China (2017YFA0303203), the National Natural Science Foundation of China (Grant Nos. 91622115, 11522432,11574217, U1732273, U1732159,

**DATA AVAILABLE**

The data that support the findings of this study are available from the corresponding author upon reasonable request.

**FIGURE CAPTIONS**

**Fig. 1. Characterization of single crystal TaSe$_3$.** **(a)** The quasi-one-dimensional crystal structure of TaSe$_3$. The blue balls refer to Ta atoms; red ones, Se atoms. **(b)** The EDS spectrum of TaSe$_3$ crystal. The insets show the EDS mapping of Ta and Se, respectively. **(c)** Raman spectrum of TaSe$_3$ crystal. **(d)** Temperature dependence of resistance of TaSe$_3$ crystal with RRR = 23.

**Fig. 2. XMR and quantum oscillations in single crystal TaSe$_3$.** **(a)** Field dependence of XMR in TaSe$_3$ with the current along the Ta-Ta chains (*b* axis) and the applied field along the [1 0 -1] direction up to 38 T at selected temperatures. **(b)** SdH patterns as a function of $1/B$ with the applied field along the *c* axis at representative temperatures. **(c)** FFT analysis of $\Delta R_{xx}$. The peaks corresponding to each oscillation mode are marked with Greek letters $\alpha$ and $\beta$. **(d)** The fits of effective mass for $\alpha$(99 T) and $\beta$(173 T) oscillation modes. **(e)** Landau fan diagram for the SdH oscillations of $\alpha$ and $\beta$ mode with *n* intercept values -0.03 and -0.02, respectively. **(f)** LK fitting of SdH pattern at 0.36 K.

**Fig. 3. First-principles calculations for TaSe$_3$.** **(a)(b)** The calculated band structure without (a) and with (b) SOC. **(c)** The schematic of the BZ of TaSe$_3$. **(d)** The calculated topological surface states on (0 1 0) plane. **(e)** The calculated WCC of six time-reversal invariant momentum planes indicates the topological invariant $Z_2$ = (1; 100). **(f)** The Fermi surfaces of TaSe$_3$. **(g)(h)** Fermi surfaces and the extremal cuts of pockets parallel with (1 0 -1) plane for $\alpha$ mode (g) and $\beta$ mode (h). In (h), the two identical electron pockets' cuts are separated by the boundary of BZ, and two parts (marked in red) make one complete cut.

**Fig. 4. Two-band model analysis for TaSe₃. (a)-(b)** Field dependence of Longitudinal resistivity $\rho_{xx}$ (a) and Hall resistivity $\rho_{xy}$ (b). **(c)-(d)** Field dependence of $\sigma_{xx}$ (c) and $\sigma_{xy}$ (d) and corresponding two-band model fitting at selected temperatures. **(e)** Temperature dependence of density of carries $n_e$, $n_h$, and the ratio $n_h/n_e$ extracted from the two-carrier model analysis of $\sigma_{xy}$. The ratio $n_h/n_e$ value stays 0.9 at *T*< 20 K. **(f)** Carrier mobility $\mu_e$, $\mu_h$ and mobility ratio $\mu_h/\mu_e$ as a function of temperature deducted from the two-carrier model analysis $\sigma_{xy}$.

**FIGURES**

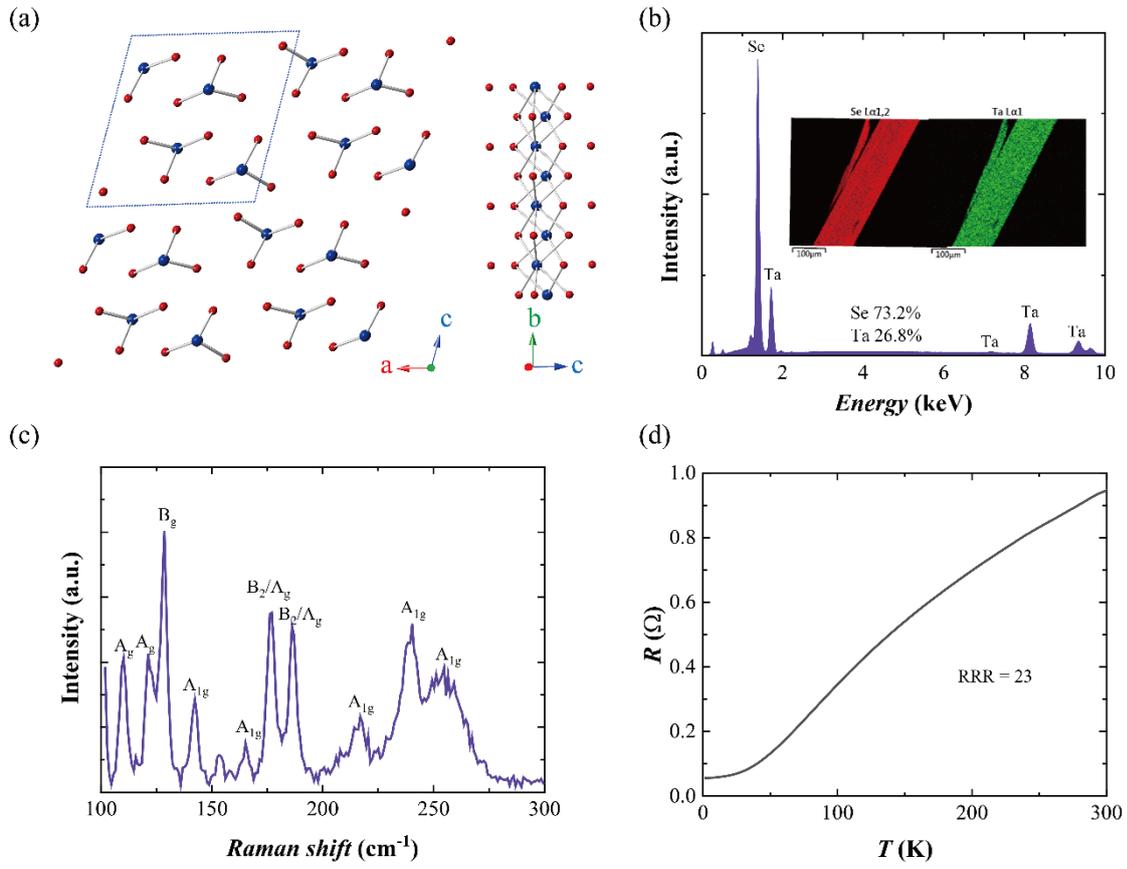

**Fig. 1**

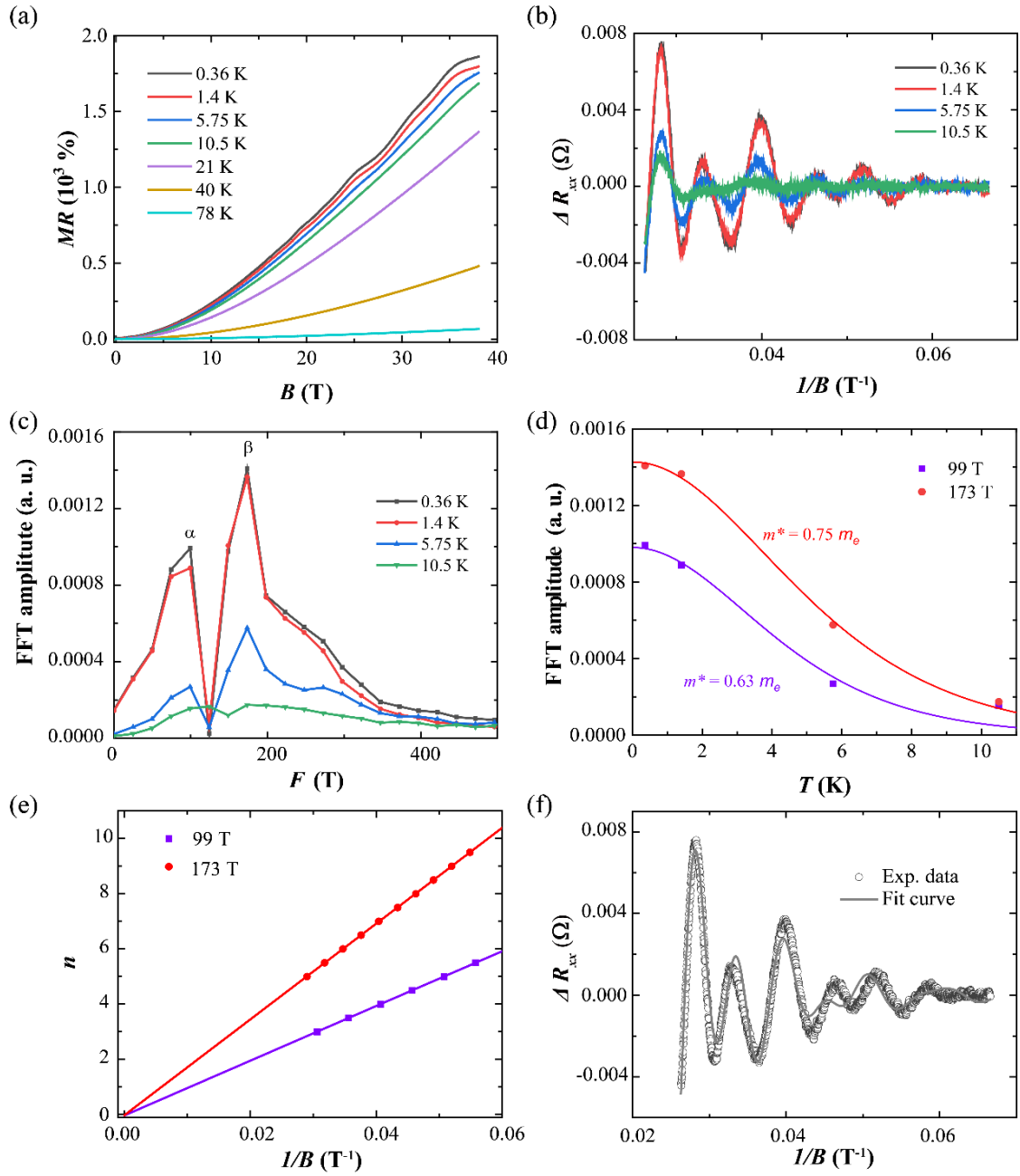

**Fig. 2**

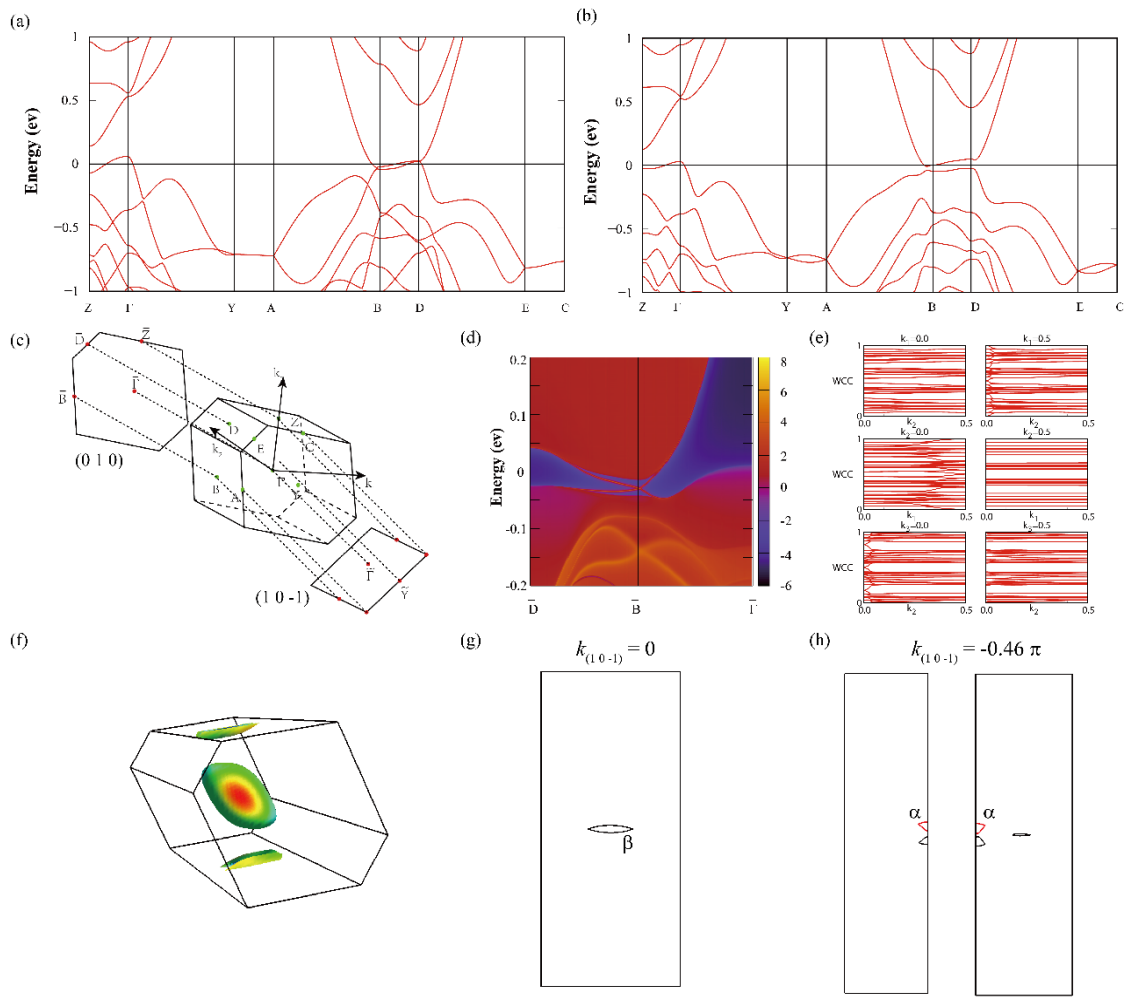

**Fig. 3**

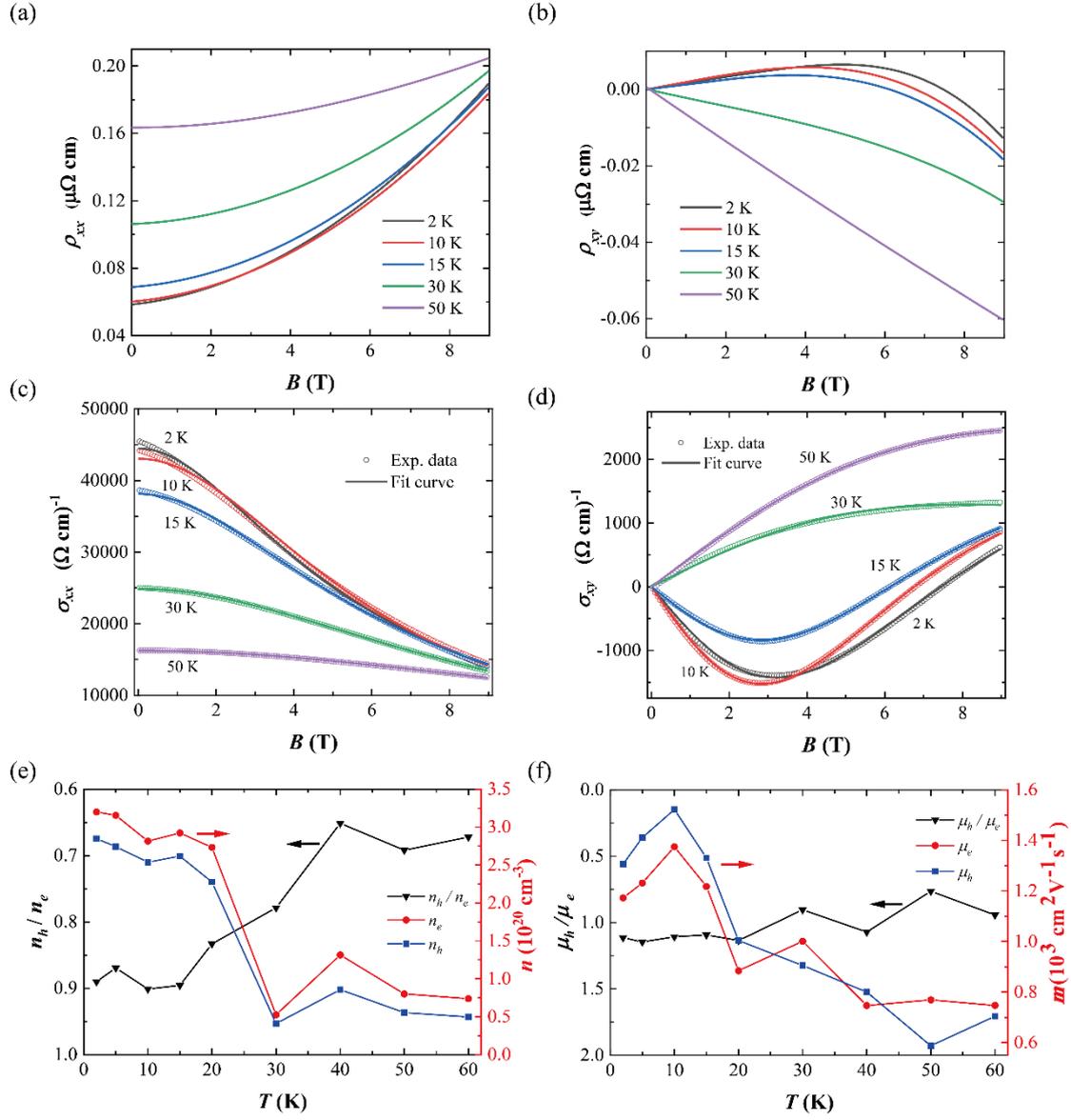

**Fig. 4**